\documentclass{PoS}
\usepackage{amsmath,amssymb}
\usepackage{dsfont}
\usepackage{cite}

\newcommand{\CP}{\ensuremath{\mathrm{CP}}}
\newcommand{\eV}{\ensuremath{\mathrm{eV}}}

\newcommand{\diag}{\ensuremath{\operatorname{diag}}}

\title{Lifting degenerate neutrino masses,
threshold corrections and maximal mixing}

\ShortTitle{Degenerate neutrinos}

\author{\speaker{Wolfgang Gregor Hollik}\thanks{Report number: TTP14-038}\\
        Institut f\"ur Theoretische Teilchenphysik\\
        Karlsruhe Institute of Technology\\
        E-mail: \email{wolfgang.hollik@kit.edu}}

\abstract{In the scenario with degenerate neutrino masses at tree-level,
  we show how threshold corrections with either non-trivial or trivial
  mixing at tree-level have the power to generate the observed
  deviations from a degenerate spectrum. Moreover, it is possible to
  also generate the mixing fully radiatively when there is trivial
  mixing at tree-level.

  We give a brief overview over the topic and discuss the outcome of
  threshold corrections for degenerate neutrino masses in a
  supersymmetric model. A detailed description can be found
  in~\cite{Hollik:2014}.}

\FullConference{Flavorful Ways to New Physics -- FWNP,\\
                28-31 October 2014\\
                Freudenstadt -- Lauterbad, Germany}

\begin{document}

\section{Introduction}
The origin of neutrino masses and mixings hints to new physics beyond
the Standard Model. Measurements of neutrino oscillations in the last
years completed our picture of neutrino flavour in a sense that now all
the mixing angles are quite precisely measured. What is still unknown is
the mass scale of the lightest neutrino, whereas the squared mass
differences are known. The masses are then to be easily calculated:
\begin{equation}
m_1 = m_0 \,,\quad
m_2 = \sqrt{m_0^2 + \Delta m_{21}^2} \,,\quad
m_3 = \sqrt{m_0^2 + \Delta m_{31}^2},
\end{equation}
where \(\Delta m_{ij}^2 = m_i^2 - m_j^2\) and we constrained ourselves
to a normal ordering in neutrino masses (\(m_3 > m_2 > m_1\)). The
precise values can be determined via a global fit on neutrino
oscillation data~\cite{Gonzalez-Garcia:2014bfa}:
\begin{equation}\begin{aligned}
\Delta m_{21}^2 &= 7.50^{+0.19}_{-0.17}\times 10^{-5}\,\eV^2 \\
\Delta m_{31}^2 &= 2.457\pm0.047\times 10^{-3}\,\eV^2.
\end{aligned}\end{equation}
In a world with \(m_0 \gg \Delta m^2\), we live in the quasi-degenerate
scenario \(m_1 \approx m_2 \approx m_3\). This situation immediately
evolves as soon as a direct mass search experiment like KATRIN reports a
positive result, which would be in this case \(m_0\approx
0.35\,\eV\)~\cite{Osipowicz:2001sq}. Cosmology also prefers degenerate
neutrino masses, where the anticipated scale is a bit below the KATRIN
limit~\cite{Battye:2013xqa} (\(m_0\approx 0.1\,\eV\)) and still
compatible with the recent \(95\,\%\) upper limit on the sum of the
three light neutrino masses obtained by the Planck
collaboration~\cite{Ade:2013zuv} (\(\sum m_\nu < 0.24\,\eV\)).

It is well known that quantum corrections become important for any type
of quasi-degenerate spectrum---renormalization group (RG)
corrections~\cite{Ellis:1999my, Casas:1999tp, Casas:1999ac, Haba:1999xz,
  Balaji:2000gd, Chankowski:2001mx, Mohapatra:2003tw, Haba:2012ar} as
well as low-energy threshold corrections~\cite{Chun:1999vb,
  Chankowski:2000fp, Chun:2001kh, Brahmachari:2003dj,
  Mohapatra:2005gs}. In general, both contributions from the RG as well
as the threshold corrections are to be added to the effective neutrino
mass matrix and change the flavour alignment as long as they are not
flavour universal:
\begin{equation}
m^\nu_{AB} = m^{(0)}_{AB} + m^{(0)}_{AC} I_{CB} + I_{AC} m^{(0)}_{CB},
\end{equation}
where \(I_{AB} = I_{AB}^\text{RG} + I_{AB}^\text{th}\) are the quantum
corrections and \(m^{(0)}_{AB}\) is the neutrino mass matrix at
tree-level. Capital indices count interaction eigenstates, where in the
following small indices represent the mass basis. By means of a unitary
transformation \(\boldsymbol{U}^{(0)}\), we arrive at the unperturbed
mass eigenbasis, where \(\boldsymbol{m}^{(0)}\) is diagonal
\begin{equation}\label{eq:masscorr}
m^\nu_{ab} = m^{(0)}_a \delta_{ab} +
\left(m^{(0)}_a + m^{(0)}_b\right) I_{ab},
\end{equation}
and \(I_{ab} = \sum_{AB} I_{AB} U^{(0)}_{Aa} U^{(0)}_{Bb}\). In the
following, we focus on the influence of the threshold corrections and
take only \(\boldsymbol{I} = \boldsymbol{I}^\text{th}\).

\section{Degenerate masses and threshold corrections}
\label{sec:thresholdcorr}
Eq.~\eqref{eq:masscorr} let us appreciate the advantage of degenerate
masses: either \(m_1 = m_2 = m_3\) and the mixing at tree-level is
trivial (\(\boldsymbol{U}^{(0)} = \mathds{1}\)), or when the masses have
opposite signs (e.g.\ \(m_1 = -m_2 = m_3\)) the corrected mass matrix
gets tremendously simplified.

The case of Majorana neutrinos does not allow to rotate away as many
\(\CP\) phases as for Dirac fermions. In general, there are two more
phases left, such that the complete diagonalisation matrix can be
written as a product of a unitary matrix with three angles and one phase
and a phase matrix: \(\boldsymbol{U}_\nu = \boldsymbol{U}^{(0)}
\boldsymbol{P}\) with \(\boldsymbol{P} = \diag{(e^{i\alpha_1},
  e^{i\alpha_2}, 1)}\). The Majorana phases \(\alpha_{1,2}\) can then be
absorbed in a redefinition of the masses instead of a redefinition of
the fields:
\[
\boldsymbol{m}^{(0)} \to \boldsymbol{P}^T \boldsymbol{U}^{(0)\,T}
        \boldsymbol{m}^{(0)} \boldsymbol{U}^{(0)} \boldsymbol{P}
        = m_0 \,\diag{(e^{-2i\alpha_1}, e^{-2i\alpha_2}, 1)}.
\]
We have chosen the phases in a particular way to have a real and
positive \(m_3\).

Under the assumption of \(\CP\) conservation in the Majorana phases, we
take \(\alpha_{1,2} \in \lbrace 0, \pm \frac{\pi}{2}\rbrace\) and get
e.g.\ \(m_1 = -m_2 = m_3\). Eq.~\eqref{eq:masscorr} simplifies to
\begin{equation}\label{eq:13corr}
  \boldsymbol{m}^\nu = m_0 \begin{pmatrix}
    1 + 2 U_{A 1} U_{\beta 1} I_{AB} & 0 &
    2 U_{A 1} U_{B 3} I_{AB} \\
    0 & - 1 -  2 U_{A 2} U_{B 2} I_{AB} & 0 \\
    2 U_{A 1} U_{B 3} I_{AB} & 0 &
    1 +  2 U_{A 3} U_{B 3} I_{AB}.
\end{pmatrix}
\end{equation}
The two-fold degeneracy reveals an initial mixing matrix at tree-level
that can be whatever it may, generated by some underlying flavour
symmetry. However, the two-fold degeneracy leaves one freedom of
rotation in the plane of degeneracy, i.e.\ the 1-3 plane:
\(\boldsymbol{U}^{(0)} \to \boldsymbol{U}^{(0)} R_{13}\) that can be
used to rotate away the off-diagonal entries in Eq.~\eqref{eq:13corr} by
requiring
\begin{equation}
\sum_{AB} U^{(0)}_{A\, 1} U^{(0)}_{B\, 3} I_{AB} = 0.
\end{equation}
Note that Majorana masses are symmetric \(m_{AB} = m_{BA}\) as are the
corrections \(I_{AB} = I_{BA}\). It is generically difficult to
accommodate for the present value of \(\sin\theta_{13}\) and the two
\(\Delta m^2\) with a minimal set of flavour-diagonal threshold
corrections. More details on an update of this scenario can be found
in~\cite{Hollik:2014}. We now shall discuss the case where there is
trivial mixing at tree-level and we generate the mixing genuinely by
threshold effects.

The case of exact degeneracy is characterized by no mixing at the
tree-level: mass and interaction eigenstates can be arbitrarily
interchanged and the fields redefined. We have three free
rotations. Therefore the observed non-trivial mixing has to be generated
via the quantum corrections only:
\begin{equation}
\boldsymbol{m}^\nu = m_0 \,\mathds{1} + m_0
\begin{pmatrix}
I_{11} & I_{12} & I_{13} \\
I_{12} & I_{22} & I_{23} \\
I_{13} & I_{23} & I_{33},
\end{pmatrix}
\end{equation}
which means that we need flavour off-diagonal corrections. How can we
constrain those? The observed neutrino mixing angles allow to first work
in some approximations and then see, whether small deviations from the
approximation help to get a better fit to data. First of all, one mixing
angle is close to maximal, \(\theta_{23} \approx \frac{\pi}{2}\), which
allows to perform the rotation in the 2-3 plane with \(I_{33} = I_{22}\)
and \(I_{23} = I_{22}\) and we obtain
\begin{equation}
  \boldsymbol{I}' = \boldsymbol{U}_{23}^T \boldsymbol{I}
  \boldsymbol{U}_{23} = \begin{pmatrix}
    I_{11} & \frac{I_{12}+I_{13}}{\sqrt{2}} & -
    \frac{I_{12}-I_{13}}{\sqrt{2}} \\ \frac{I_{12}+I_{13}}{\sqrt{2}} & 2
    I_{22} & 0 \\ -\frac{I_{12}-I_{13}}{\sqrt{2}} & 0 & 0
  \end{pmatrix},
\end{equation}
which, together with the observation that one mixing angle is
small\footnote{Recent data tell us rather \(\theta_{13} \approx
  9^\circ\).For the moment, we stick to \(\theta_{13} = 0^\circ\).}
\(\theta_{13} \approx 0\) allows for the approximation \(I_{13} \approx
I_{12}\) and the last rotation is done with
\begin{equation}
\theta_{12} \approx \frac{1}{2} \arctan\left(\frac{2\sqrt{2} I_{12}}{2
    I_{22} - I_{11}}\right).
\end{equation}
In this way, there is one free mixing angle \(\theta_{12}\) and the two
\(\Delta m^2\) to be determined by three corrections \(I_{11}\),
\(I_{22}\) and \(I_{12}\), which is good in the approximation but
obviously not sufficient to fit all the data. The deviation of
\(\theta_{23}\) from \(\frac{\pi}{2}\) can be obtained by a splitting
\(I_{33} = I_{22} + \varepsilon\) and \(\theta_{13} \neq 0\) with
\(I_{13} = I_{12} + \delta\). We then have five parameters \(I_{11},
I_{12}, I_{22}, \varepsilon\) and \(\delta\) to determine three mixing
angles \(\theta_{12}, \theta_{13}, \theta_{23}\) and two mass splittings
\(\Delta m_{21}^2\) and \(\Delta m_{31}^2\).
\begin{align}
&\begin{pmatrix}
m_0 &&\\ &\sqrt{m_0^2+\Delta m_{21}^2}& \\ &&\sqrt{m_0^2+\Delta m_{31}^2}
\end{pmatrix}
=\\&\qquad\qquad\qquad\qquad\qquad m\, \boldsymbol{U}(\theta_{12},\theta_{13},\theta_{23})^T
\begin{pmatrix}
1 + I_{11} & I_{12} & I_{12} + \delta \\
I_{12} & 1 + I_{22} & I_{22} \\
I_{12} + \delta & I_{22} & 1 + I_{22} + \varepsilon
\end{pmatrix}
\boldsymbol{U}(\theta_{12},\theta_{13},\theta_{23}),
\nonumber
\end{align}
the degenerate mass parameter \(m\) can be identified with \(m_0\). Using the possible KATRIN discovery, \(m_0 = 0.35\,\eV\), we get
\begin{equation}\label{eq:result}
\boldsymbol{I} = \begin{pmatrix}
0.976 & 1.03 & 1.05 \\
1.03 & 4.75 & 4.75 \\
1.05 & 4.75 & 6.74
  \end{pmatrix} \times 10^{-3}.
\end{equation}

\section{Threshold corrections in the \(\mathbf{\nu}\)MSSM}
The generic threshold corrections given in the previous section have the
property of large flavour off-diagonal contributions. One viable model
to produce such corrections is the Minimal Supersymmetric Standard Model
with right-handed neutrinos (\(\nu\)MSSM) that generate small neutrino
masses via a see-saw mechanism.

\begin{figure}[tb]
\begin{minipage}{.5\textwidth}
\includegraphics[width=\textwidth]{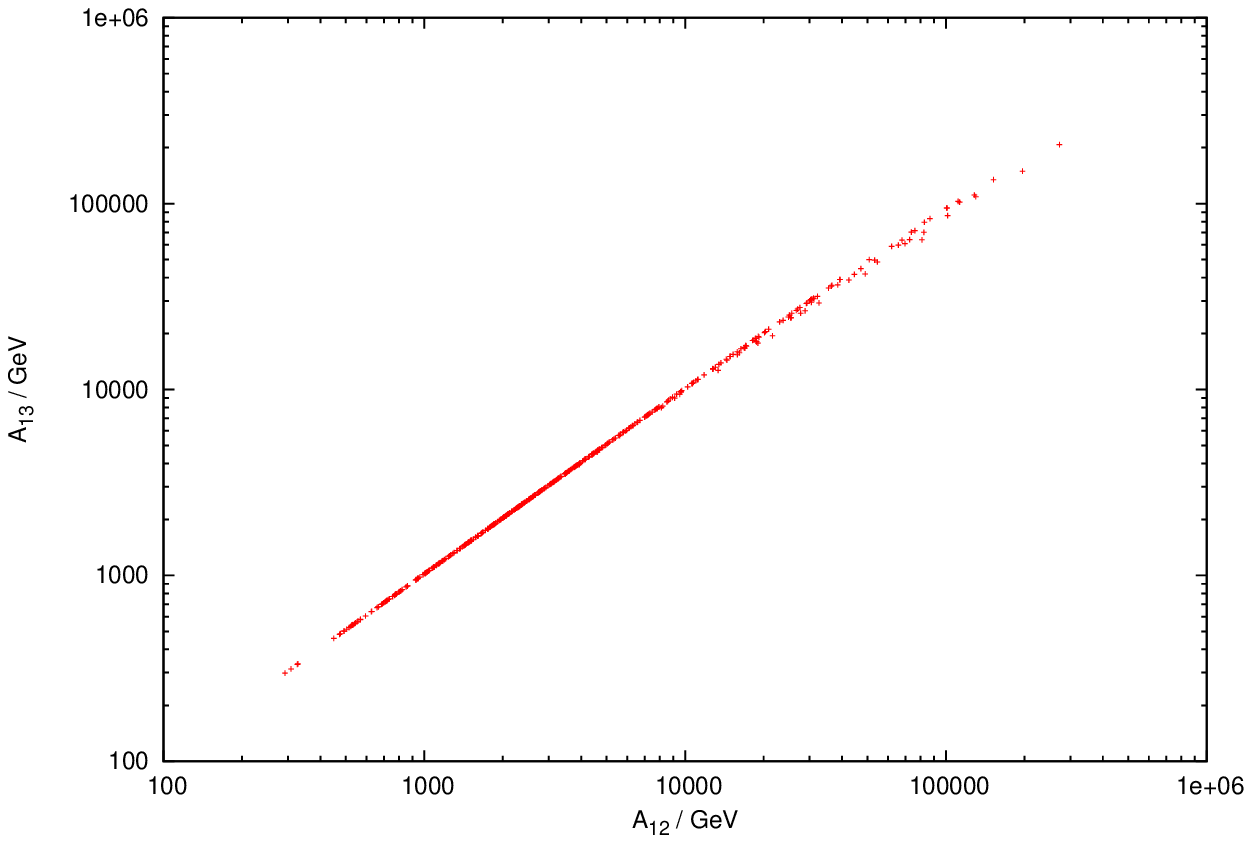}
\end{minipage}%
\begin{minipage}{.5\textwidth}
\includegraphics[width=\textwidth]{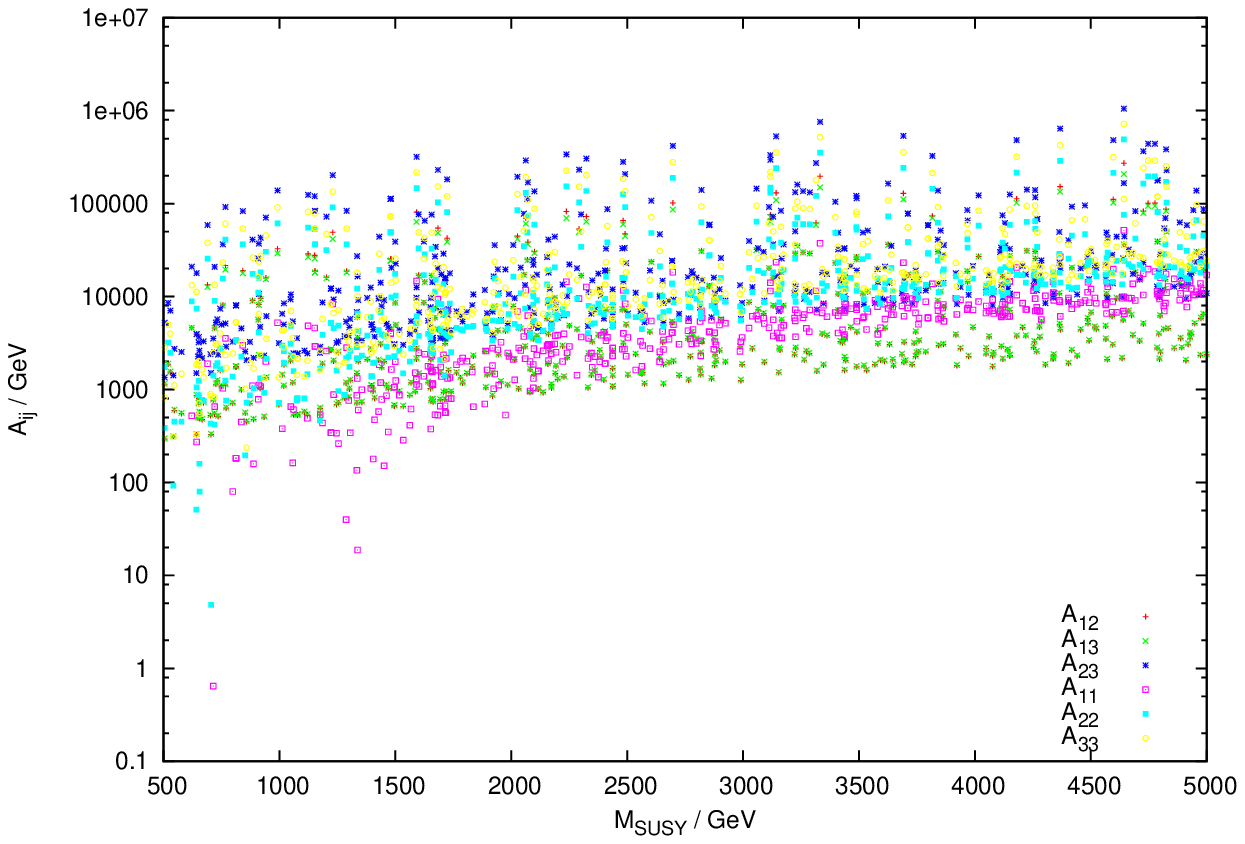}
\end{minipage}%
\caption{Results of the fit for a scan over several free parameters
  (details in~\cite{Hollik:2014}). The left plot shows the correlation
  between \(A^\nu_{12}\) and \(A^\nu_{13}\) which similar to the generic
  results of Sec.~2. On the right side the typical size of
  \(A^\nu_{ij}\) is shown with respect to the soft breaking mass
  \(M_\mathrm{SUSY}\).}
\label{fig:results}
\end{figure}

The superpotential of the model is given by
\begin{equation}
\mathcal{W} \supset \mu H_1 \cdot H_2
 + Y^\nu_{ij}\; H_2 \cdot L_{L,i} N_{R,j}
 - Y^\ell_{ij}\; H_1 \cdot L_{L,i} E_{R,j}
 + \frac{1}{2} M^R_{ij} N_{R,i} N_{R,j},
\end{equation}
and the soft breaking terms
\begin{equation}
V_\mathrm{soft}^{\tilde\nu} =
\left(\boldsymbol{m}_{\tilde L}^2\right)_{ij} \tilde{\nu}_{L,i}^*\tilde\nu_{L,j}
+
\left(\boldsymbol{m}_{\tilde R}^2\right)_{ij} \tilde\nu_{R,i}\tilde{\nu}^*_{R,j}
+
\left(
A^\nu_{ij}\; h_2^0\, \tilde\nu_{L,i}\tilde{\nu}_{R,j}^*
+
\left(\boldsymbol{B}^2\right)_{ij} \tilde{\nu}_{R,i}^*\tilde{\nu}_{R,j}^*
+ \text{h.~c.\ }
\right),
\end{equation}
that have the power to introduce flavour-changing self-energies. We take
the soft masses universal, \(\boldsymbol{m}_{\tilde L}^2 =
\boldsymbol{m}_{\tilde R}^2 = M_\mathrm{SUSY} \mathds{1}\), and
determine values of \(A^\nu_{ij}\) to give corrections like
Eq.~\eqref{eq:result}. The results are shown in Fig.~\ref{fig:results},
details on the calculation of threshold corrections in the \(\nu\)MSSM
are given in~\cite{Hollik:2014}.

\section{Conclusions}
We have discussed degenerate neutrino masses at tree-level and
investigated threshold corrections to two different mass patterns. In
the scenario with one mass of a different sign, the minimal set of
threshold corrections cannot produce the necessary deviation from
degeneracy for the masses. On the other hand, generic threshold
corrections can simultaneously generate all three mixing angles and the
mass differences in the case of exactly degenerate masses and trivial
mixing at tree-level. These findings have been applied to a
supersymmetric model where the soft breaking terms carry the flavour
information.

\section*{Acknowledgements}
This work was supported by the GRK 1694 \emph{``Elementarteilchenphysik
  bei h\"ochster Energie und h\"ochster Pr\"azision''}. I appreciate
valuable discussions with S.~Pokorski about that topic. The workshop
``Flavorful Ways to New Physics'' was supported by the Karlsruhe House
of Young Scientists (KHYS), the Karlsruhe School of Elementary Particle and
Astroparticle Physics (KSETA), the KIT Center for Elementary Particle
and Astroparticle Physics (KCETA) and the Graduiertenkolleg GRK 1694 as
well as BlueYonder.

\end{document}